\documentstyle[a4,12pt,epsf]{article}
\def\be{\begin{equation}}
\def\ee{\end{equation}}
\def\bq{\begin{eqnarray}}
\def\eq{\end{eqnarray}}
\newcommand{\ra}{\rightarrow}
\newcommand{\g}{\gamma}
\newcommand{\vp}{\varphi}
\newcommand{\al}{\alpha}

\begin{document}
\thispagestyle{empty}
\setcounter{page}{0}
\setcounter{page}{0}
\begin{flushright}
MPI-PhT/95-19\\
LMU 05/95\\
March 1995
\end{flushright}
\vspace*{\fill}
\begin{center}
{\Large\bf Heavy Meson Exclusive Decays in the Framework of QCD Sum Rules}
$^{1,*}$\\
\vspace{2em}
\large
A. Khodjamirian$^{a,\dagger}$ and R. R\"uckl$^{a,b}$\\
\vspace{2em}
{$^a$ \small Sektion Physik der Universit\"at
M\"unchen, D-80333 M\"unchen, Germany }\\
{$^b$ \small Max-Planck-Institut f\"ur Physik,
Werner-Heisenberg-Institut,
D-80805 M\"unchen, Germany}\\

\end{center}
\vspace*{\fill}

\begin{abstract}
We discuss applications of QCD sum rules on the light-cone
to the form factors of
the exclusive transitions $B \rightarrow \pi$ and $D\rightarrow \pi$,
and to the $B^*B \pi$ and $D^*D \pi$ coupling constants.
In the light of our results we examine the pole dominance model for
these form factors. A first estimate is given on the nonfactorizable
amplitude of the decay $ B \ra J/\psi K$.
\end{abstract}

\vspace*{\fill}

\begin{flushleft}

\noindent$^\dagger$on leave from Yerevan Physics Institute,
375036 Yerevan, Armenia \\
\noindent$^1${\small supported by the German Federal Ministry for
Research and Technology (BMFT) under contract No. 05 6MU93P
and by the EC-program HCM under contract No. C.E.E. CT93-0132}
\\
\noindent$^*${\it presented by R. R\"uckl at the
138.WE-Heraeus-Seminar on Heavy Quark Physics, Bad Honnef, Germany,
December 14-16,1994}
\baselineskip=16pt
\end{flushleft}

\newpage
\section{Introduction}

The reliable extraction
of fundamental parameters from data on heavy flavoured hadrons
is an important theoretical task.
While the inclusive $B$ and $D$ decays appear to
be the cleanest reactions theoretically, exclusive decays are
experimentally often more favourable. However, for the
interpretation of exclusive measurements one needs
an accurate knowledge of
decay constants, form factors and other hadronic
matrix elements. Among the existing approaches,
QCD sum rules \cite{SVZ} have proved
to be particularly powerful in obtaining reliable estimates. In this
report, we discuss applications of the sum rule method to
form factors of the transitions $D \ra \pi$ and $B\ra \pi$,
to the $D^*D \pi$ and $B^*B \pi$ coupling constants,
and to the nonfactorizable
amplitude of the decay $ B \ra J/\psi K$. From
a more technical point of view, our calculations aim
at developing alternative variants of sum rules which
avoid  some of the problems inherent in the more familiar
original version.

As explained in Section 2, the so-called light-cone sum rules
provide a very economical way to obtain $B$ and $D$
form factors and couplings. In this variant, the ideas
of duality and matching
between parton and hadron descriptions intrinsic to
QCD sum rules are combined with the operator
product expansion (OPE) techniques used to study hard exclusive
processes in QCD \cite{BLreport,CZreport}.
Using these results , we then examine the pole dominance
model for form factors in Section 3. Finally, in Section 4 we
describe an attempt to estimate weak amplitudes beyond the usual
factorization approximation considering
the decay mode $ B \rightarrow J/\psi K $  as a prototype example
and employing conventional sum rule methods.

\section{Transition form factors and hadronic couplings}

\subsection{ QCD sum rules on the light-cone }

In contrast to the conventional sum rules based on
the Wilson OPE of the T-product of currents at small
distances, one may consider expansions near the light-cone
in terms of nonlocal operators, the matrix elements of which are
given by  hadron
wave functions of increasing twist. As one advantage, this
formulation allows to incorporate additional
information about the Euclidean asymptotics of correlation functions
in QCD  for arbitrary external momenta.

For definiteness, we focus on the  correlation function which will
later be used to evaluate the form factor $D\ra\pi$ and the
$D^{*}D \pi$ coupling:

$$
F_\mu (p,q)=
i\int d^4xe^{ipx} \langle \pi^-(q)|T\{ \bar{d}(x)\g_\mu c(x),
\bar{c}(0)i\g_5 u(0)\} |0\rangle
$$
\be
=F(p^2,(p+q)^2)q_\mu + \tilde{F}(p^2,(p+q)^2)p_\mu ~.
\label{19}
\ee
With the pion on mass-shell, $q^2=m_\pi^2 $, the
correlation function (\ref{19})
depends on two invariants, $p^2$ and $(p+q)^2$.
We set $m_\pi=0$ everywhere.

In the Euclidean region
where both $p^2$ and $(p+q)^2$ are negative
and large, the charm quark is far off-shell.
Substituting, as a first approximation,
the free $c$-quark propagator
\bq
\langle 0|T\{c(x)\bar{c}(0)\}|0\rangle = i\hat{S}_c^0(x) =
\int \frac{d^4k}{(2\pi)^4i}e^{-ikx}
\frac{\not\!k+m_c}{m_c^2-k^2}
\label{prop}
\eq
into eq. (\ref{19}) one readily obtains
\bq
F_\mu(p,q)&=&i\int \frac{
d^4x\,d^4k}{(2\pi )^4(m_c^2-k^2)}
e^{i(p-k)x}\left(m_c
\langle \pi (q)|\bar{d}(x)\g_\mu\g_5u(0)|0\rangle \right.
\nonumber
\\
&&{}+\left.
k^\nu \langle\pi(q) |\bar{d}(x)\g_\mu\g_\nu\g_5u(0)|0\rangle\right)~.
\label{23}
\eq
 This contribution is depicted diagramatically in Fig.~1a.

Short-distance expansion of the first matrix element
of eq. (\ref{23}) in terms of local operators,
\be
\bar{d}(x)\g_\mu\g_5u(0)=\sum_n\frac{1}{n!}
\bar{d}(0)(\stackrel{\leftarrow}{D}\cdot x)^n\g_\mu\g_5 u(0)~,
\label{expan}
\ee
and integration over $x$ and $k$ yield
\bq
F_\mu(p,q)&=&i\frac{m_c}{m_c^2-p^2}
\sum_{n=0}^\infty \frac{(2p \cdot q)^n}{(m_c^2-p^2)^n}M_n q_\mu ~,
\label{expans}
\eq
where
$$\langle\pi(q) |\bar{d}\stackrel{\leftarrow}{D}_{\alpha_1}
\stackrel{\leftarrow}{D}_{\alpha_2}...
\stackrel{\leftarrow}{D}_{\alpha_n}\g_\mu\g_5 u  |0\rangle
=(i)^n q_\mu q_{\alpha_1} q_{\alpha_2}...q_{\alpha_n}M_n + ...~ $$
has been used, $D$ being the covariant derivative.
One now encounters the following problem. If
the ratio
\bq
\tilde{\xi}=2(p \cdot q)/(m_c^2-p^2)= ((p+q)^2-p^2)/(m_c^2-p^2)
\label{ksi}
\eq
is finite one must keep an {\em infinite} series of matrix elements  of
local operators in eq. (\ref{expans}). All of them give
contributions of the order $1/(m_c^2-p^2)$ in the heavy quark
propagator, differing only by powers of the dimensionless parameter
$ \tilde{\xi}$.
Therefore, short-distance expansion of eq. (\ref{23}) is
useful {\em only} if $ \tilde{\xi} \rightarrow 0$,
i.e. for $p^2 \simeq (p+q)^2$
or, equivalently, $q \simeq 0$. In this case, the series in
eq. (\ref{expans}) can be truncated after a few terms
involving only a small
number of unknown matrix elements $M_n$. However, for general momenta
with
$p^2 \neq (p+q)^2$ one has to sum up the infinite series of matrix
elements of local operators in some way.

This formidable task can be solved by using
techniques developed for hard exclusive processes in QCD
\cite{BLreport,CZreport}.
Returning to the initial expression (\ref{19})
for the correlation
function one expands the $T$-product of currents
near the light-cone $x^2=0$. In a first step this leads to the same
approximation
(\ref{23}) involving
vacuum-to-pion transition matrix elements of nonlocal operators
composed of light quark fields at light-like separation.
These matrix elements are expanded in $x$
and at $ x^2 \simeq 0 $ reexpressed in terms of  pion wave functions
with given twist. For the present discussion it
is again sufficient to focus on the  first term in eq. (\ref{23})
proportional to $m_c$.
In leading twist one has
\bq
\langle\pi(q)|\bar{d}(x)\g_\mu\g_5u(0)|0\rangle=
-iq_\mu f_\pi\int_0^1du\,e^{iuqx}\vp_\pi (u)  ~,
\label{pionwf}
\eq
where the wave function  $\vp_\pi$ represents the distribution in the
fraction $u$
of the light-cone momentum $q_0 +q_3 $ of the pion carried by
a constituent quark.
Substituting eq. (\ref{pionwf}) in eq. (\ref{23})
and integrating over $x$ and $k$ one finds for the invariant function
$F$ defined in eq. (\ref{19}):
\be
F(p^2,(p+q)^2)=m_cf_\pi\int_0^1\frac{du ~\vp_\pi(u) }{m_c^2-(p+uq)^2}
+...~,
\label{Fzeroth}
\ee
where the ellipses represent contributions of higher twists and
multicomponent wave functions. The leading three-particle wave function
enters in connection with gluon emission
by the heavy quark line as shown in Fig. 1b.
This contribution is included in the calculations
of refs. \cite{BKR,BBKR} as well as two-particle wave functions up
to twist 4. The calculation of perturbative $O(\alpha_s)$ corrections
indicated in Fig. 1c and 1d is in progress.

Comparing eqs. (\ref{expans}) and (\ref{Fzeroth}) one sees that the
infinite series of matrix elements of local operators
encountered before in eq. (\ref{expans})  is effectively replaced
by hadronic wave functions. These universal functions
describe the long-distance dynamics similarly as the universal
vacuum condensates appearing in the more familiar sum rule variant
based on short-distance expansion. The universality property is essential
for the light-cone approach.
\vspace*{3.5cm}
\begin{center}
\begin{minipage}[t]{7.8cm} {
\begin{center}
\hspace{-3.2cm}
\mbox{
\epsfysize=10cm
\epsffile[0 0 500 500]{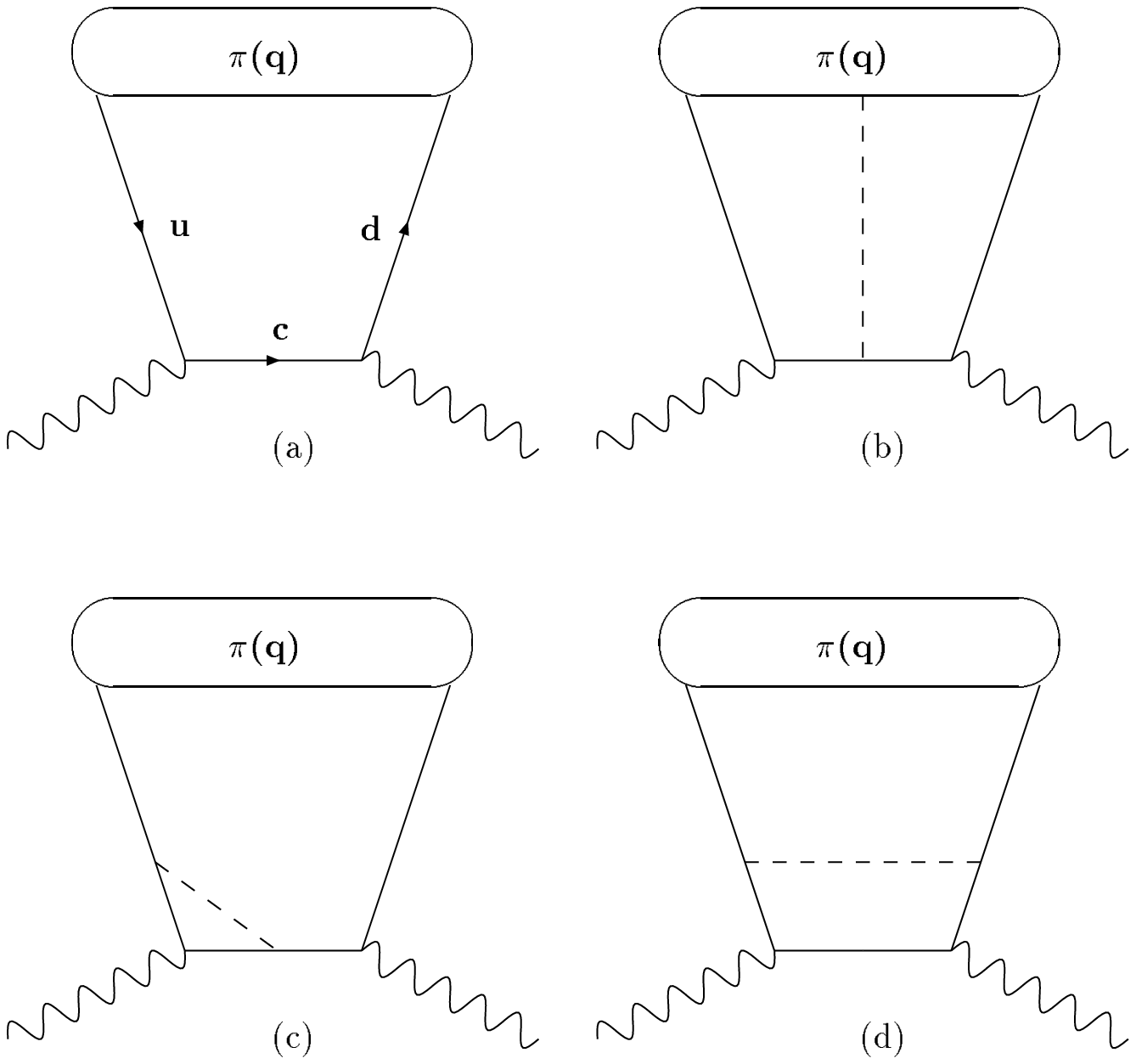}
}
\end{center}
}\end{minipage}
\end{center}
\vspace{-3.5cm}
{\small\bf Fig.~1. }{\small\it
QCD diagrams contributing to the correlation function
(1) and involving (a) quark-antiquark light-cone wave functions, (b)
three-particle quark-antiquark-gluon wave functions,
(c) and (d) perturbative $O(\alpha_s)$ corrections.
Solid lines represent quarks, dashed lines gluons,
wavy lines are external currents. }


\subsection{ $D \ra \pi$ and $B \ra \pi$ form factors}
The light-cone sum rule for the form factor  $f^+_D(p^2)$ entering the
transition amplitude
\be
\langle \pi(q) \mid \bar d\g_\mu c \mid D(p+q) \rangle =
2f^+_D(p^2)q_\mu + ( f^+_D(p^2) +f^-_D(p^2))p_\mu
\label{formfdef}
\ee
is obtained by matching the expression  (\ref{Fzeroth})
for the invariant amplitude $F(p^2, (p+q)^2)$
in terms of pion wave functions
with the hadronic representation
\be
F(p^2, (p+q)^2) = \frac{2m_D^2f_Df^+_D(p^2)}{m_c(m_D^2-(p+q)^2)}
+ \int_{s_0}^{\infty} \frac{\rho^h(p^2,s)ds}{s- (p+q)^2} ~.
\label{physpart}
\ee
In the above, the pole term is due to the ground state in the heavy
channel, while the excited and continuum states
are taken into account by the dispersion integral with the
effective threshold $s_0$. Invoking semilocal duality,
the latter contributions are cancelled against the
corresponding piece of the dispersion integral representation of the
QCD result on the l.h.s. of eq.
(\ref{physpart}). After Borel transformation
in the variable $(p+q)^2$, i.e. after applying the operator
\be
{\cal B}_{M^2}f(Q^2)=lim_{Q^2,n \ra \infty , Q^2/n=M^2 }
\frac{(Q^2)^{(n+1)}}{n!}\left( -\frac{d}{dQ^2}\right)^n f(Q^2)
\equiv f(M^2)
\label{B}
\ee
to eq. (\ref{physpart}) where $M^2$ is called the Borel parameter,
one finds the following sum rule:
$$
f_Df^+_D( p^2)= \frac{f_\pi m_c^2}{2m_D^2}
\Bigg \{\int_\Delta^1\frac{du}{u}
\exp\left[\frac{m_D^2}{M^2}-\frac{m_c^2-p^2(1-u)}{uM^2}\right]
\Phi_2(u,M^2,p^2)
$$
$$
-\int_0^1\!\!u du\!\int_0^1 d\al_1\int_0^{1-\al_1} d\al_2
\frac{
\Theta( \alpha_1+u\alpha_2-\Delta)}{(\alpha_1+u\alpha_2)^2}
$$
\be
\times \exp\!\left[\frac{m_D^2}{M^2}-\frac{m_c^2-p^2
(1-\alpha_1-u\alpha_2)}{(\alpha_1+
u\alpha_2)M^2}\right]\!\Phi_3(u,M^2,p^2) \Bigg\}\nonumber~,
\label{formSR}
\ee
where  $\Delta =(m_c^2-p^2)/(s_0^{(c)}-p^2)$ and
\be
\Phi_2 = \vp_\pi(u) +
\frac{\mu_\pi}{m_c}\Bigg[u \vp _{p}(u)
+ \frac16 \vp_{\sigma }(u)
\left(2 + \frac{m_c^2+p^2}{uM^2}\right)\Bigg]+...~,
\label{phi2}
\ee
\be
\Phi_3= \frac{2f_{3\pi}}{f_{\pi}m_c}
\varphi_{3\pi}(\al_1,1-\al_1-\al_2,\al_2)
\left[1-\frac{ m^2_c -p^2 }{(\alpha_1+u\alpha_2)M^2}\right] +...~.
\label{phi3}
\ee
Here, $\varphi_p$, $\varphi_\sigma$,
and $\varphi_{3\pi}$ are twist-3
pion wave functions. The ellipses denote contributions of higher
twist. The contributions of twist 4 are given explicitly
in refs. $^{4,5}$.
The analogous sum rule  for the  $B \ra \pi$ form factor is obtained
from the above by formally changing $c \ra b $ and $D \ra \bar B $.

The numerical values to be substituted
for $m_c$, $f_D$ and $s_0$ are interrelated by the
QCD sum rule for the two-point correlation
function  $\langle$ 0 $\mid T \{ j_5(x),j^+_5(0)\}\mid$ 0 $\rangle$,
$j_5= \bar{c}i\gamma_5 u$.
We use the set [$f_D = 170 \pm 10$ MeV , $ m_c=1.3$ GeV,
$s_0^{(c)}=6$ GeV$^2$] which satisfies this two-point sum rule
without $O(\alpha_s)$ corrections in consistency with the
neglect of $O(\alpha_s)$ corrections in the sum rule for $f_Df_D^+$.
The uncertainty quoted for $f_D$ corresponds to the
variation with the Borel parameter $M^2$ within the appropriate range
of $M^2$.
A similar interrelation exists for $m_b$, $f_B$ and $s^{(b)}_0$,
where
the analogous two-point sum rule
yields the set of values [$f_B = $140 MeV, $m_b=$4.7 GeV,
$s_0^{(b)}$= 35 GeV$^2$]. The variation of $f_B$ with $M^2$ is negligible.

For the pion wave functions we use the parametrization  suggested
in ref. \cite{BF1}. Arguments for this choice are given in ref.
\cite{BBKR}.
The maximum momentum transfer $p^2$ at which the sum rule
(\ref{formSR}) is
applicable is estimated to be about 1 GeV$^2$ for $D$ mesons
and 15 GeV$^2$ for $B$ mesons. The resulting form factors
$f^+_D( p^2)$ and $f^+_B( p^2)$
are plotted in Fig. 2. The
dependence of eq. (\ref{formSR})
on the Borel parameter $M^2$ is rather weak in
the range where the twist-4 and the continuum contributions are
less than 10\% and 30\%, respectively \cite{BKR}. For definiteness,
we have taken $M^2=4$ GeV$^2$ for the $D\ra\pi$ form factor
and $M^2=10$ GeV$^2$ for the $B \ra \pi$ form factor plotted
in Fig. 2.
\newpage
\vspace*{4.8cm}
\begin{center}
\begin{minipage}[t]{7.8cm} {
\begin{center}
\hspace{-3.2cm}
\mbox{
\epsfysize=10cm
\epsffile[0 0 500 500]{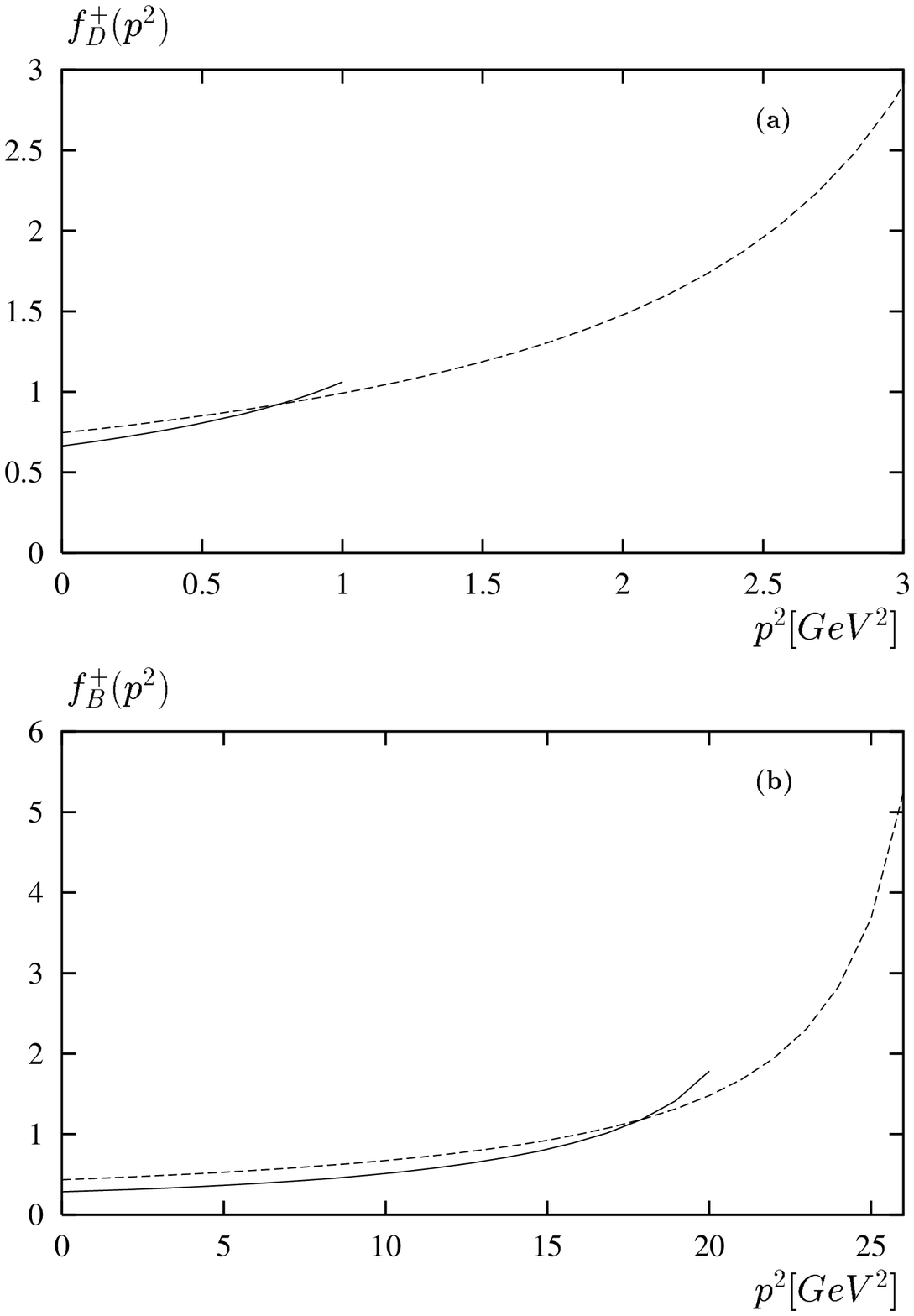}
}
\end{center}
}\end{minipage}
\end{center}
\vspace{-2.5cm}
{\small\bf Fig.~2. }{\small\it
The form factors for the transitions
(a) $D \rightarrow \pi$  and (b) $B \rightarrow \pi$
as predicted by the light-cone sum rule (solid lines)
in comparison to the single-pole approximation
(dashed lines) with the normalization fixed by the coupling
constants $g_{D^*D\pi }$ and $g_{B^*B \pi }$, respectively,
determined from the analogous sum rules.}
\\

\subsection{$D^*D\pi$ and $B^*B\pi$ couplings}

Next we sketch how the relation (\ref{Fzeroth}) can be turned into a
sum rule for the coupling constant $g_{D^*D\pi}$.
The key idea is to write a double dispersion integral for the
invariant function $F$:
\bq
F(p^2,(p+q)^2)=\frac{m_D^2m_{D^*}f_Df_{D^*}g_{D^*D\pi}}{m_c
(p^2-m_{D^*}^2)((p+q)^2-m_D^2)}
\nonumber
\\
+\int\frac{\rho^h(s_1,s_2)ds_1ds_2}{(s_1-p^2)(s_2-(p+q)^2)}
\nonumber
\\
+\int\frac{\rho ^h _1(s_1)ds_1}{s_1-p^2}
+\int\frac{\rho ^h _2(s_2)ds_2}{s_2-(p+q)^2}~.
\label{221}
\eq
Here, the first term arises from the ground state contribution
and contains the $D^*D\pi$ coupling defined by the on-shell
matrix element
\be
\langle D^{*+}(p)\pi^-(q)\mid D^0(p+q)\rangle =
-g_{D^*D\pi}q_\mu \epsilon ^{\mu} ,
\label{99}
\ee
while  the spectral function $\rho^h(s_1,s_2)$
represents higher resonances and continuum states in the
$D^*$ and $D$ channels. The additional single dispersion integrals
are due to necessary subtractions. Then, considering
$p^2$ and $(p+q)^2$ as independent
variables and
applying the Borel operator (\ref{B}) to eq. (\ref{221}) with respect to
both $p^2$ and $(p+q)^2$, we obtain
\bq
F(M_1^2,M_2^2)&\equiv&
{\cal B}_{M_1^2}{\cal B}_{M_2^2}F(p^2,(p+q)^2)=
\frac{m_D^2m_{D^*}f_Df_{D^*}g_{D^*D\pi} }{m_c}
e^{-\frac{m_{D^*}^2}{M_1^2}-\frac{m_{D}^2}{M_2^2}}
\nonumber
\\
&+&\int e^{-\frac{s_1}{M_1^2}-\frac{s_2}{M_2^2}}
\rho^h(s_1,s_2)ds_1ds_2  ~,
\label{222}
\eq
where  $M_1^2$ and $M_2^2$  are the Borel parameters associated with
$p^2$ and $(p+q)^2$, respectively.
Note that
contributions from heavier states are now exponentially suppressed
by factors
$\exp\{-\frac{s_{1,2}^2-m_{D^*,D}^2}{M_{1,2}^2} \}$ as desired, while
the subtraction terms depending only on one of the
variables, $p^2$ or $(p+q)^2$, vanish.

Applying the same transformation to
the expression (\ref{Fzeroth}) and equating the result with eq.
(\ref{222})
we end up with the  sum rule
\begin{equation}
\frac{m_D^2m_{D^*}f_Df_{D^*}}{m_c} \cdot g_{D^*D\pi}
= m_c f_\pi~\vp_\pi(u_0)M^2
\exp \left[ \frac{m^2_{D^\ast}-m_c^2}{M_1^2} +
\frac{m^2_{D}-m_c^2}{M_2^2}\right] +\ldots
\label{SR0}
\end{equation}
where $u_0=\frac{M_1^2}{M^2_1+M_2^2}$ and
$M^2=\frac{M_1^2M^2_2}{M^2_1+M_2^2}$.
The ellipses refer to higher-twist
and gluonic contributions. The contributions from higher states
are again subtracted invoking semilocal duality as discussed in detail
in ref. \cite{BBKR}.

Since $M_1^2$ and $M_2^2$ are expected to be quite similar in
magnitude,
the coupling constant $g_{D^*D\pi}$ is determined by the value of the
pion wave function at $u\simeq 1/2$,
that is by the probability for the quark and
the antiquark to carry equal momentum fractions in the pion.
This interesting feature is shared by the sum rules for many other
important hadronic couplings involving the pion.
As already pointed out, the quantity $\vp_\pi(1/2)$  is considered to be
a universal nonperturbative parameter, similar to
quark and gluon condensates in the standard approach. It
may be determined from suitable sum rules in which the
phenomenological part is known experimentally. We take the
value $\vp_\pi(1/2) = $1.2$\pm$0.2
obtained from the light-cone sum rule for the
pion-nucleon coupling \cite{BF1}. For the remaining parameters
we use the same input values as in the calculation of the form factor
$f_D^+$ in Section 2.2.
In addition, we take
$f_{D^*}$ = 240 $\pm$ 20 MeV as determined from the corresponding
two-point sum rule. With this choice, we obtain
\be
g_{D^*D\pi}= 12.5\pm 1.0~.
\label{gDDpi}
\ee
The uncertainty indicates the variation of $g_{D^*D\pi}$
in the interval
2 GeV$^2< M^2<$ 4 GeV$^2$,
where the higher state contributions are less than 30\%
and the  twist-4 corrections do not exceed 10\% .
The sensitivity to the effective threshold $s_0$ is reasonably
small. For example, variation of $s_0^{(c)}$
between  5 and 7 GeV$^2$, while
all other parameters are kept fixed, leads to a total variation of
the coupling $g_{D^*D\pi}$ by
less than 5\%. The above prediction can be directly
tested experimentally in the decay $D^* \rightarrow D \pi $.
Eq. (\ref{gDDpi}) implies the decay width
$
\Gamma( D^{*+} \rightarrow D^0 \pi^+)~ =
 ~32 \pm 5 \,\mbox{\rm keV}
$,
which is well below the current experimental upper limit\cite{ACCMOR,CLEO1}
$
\Gamma( D^{*+} \rightarrow D^0 \pi^+)~ < ~ 89 ~\mbox{\rm keV}
$.

The sum rule for $g_{D^*D\pi}$ given in eq. (\ref{SR0})
is easily converted into a sum rule for
the coupling $g_{B^*B\pi}= g_{\bar B^{*0}B^-\pi^+}$
by replacing  $c$ with  $b$, $D$ with
$\bar B$, and $D^*$ with $\bar B^*$. Using $f_{B^*}= 160$ MeV
in addition to the $B-$channel parameters specified in Section 2.2 and
confining oneself to the
corresponding fiducial interval 6 GeV$^2$ $< M^2 <$ 12 GeV$^2$, one
finds
\be
g_{B^*B\pi}=29\pm 3~.
\label{gBBpi}
\ee
If the threshold
$s_0^{(b)}$ is varied
between 34 and 36 GeV$^2$, this value changes by 5\%.

The dependence on the pion wave function disappears in
the limit $q\rightarrow0$ as can be seen from
eq. (\ref{Fzeroth}) because of the normalization condition
$ \int^1_0 du \varphi_\pi(u) = 1$. This is just the limit where the
correlation
function (\ref{19}) can be treated in short-distance expansion.
The condition $q \simeq 0$
is also implicitly assumed in refs.
\cite{EK85,Cetc94} where the correlation function (\ref{19})
is calculated using the external field method, or equivalently
the soft-pion approximation.
Our more general calculation \cite{BBKR} confirms the result
of ref. \cite{Cetc94}.
\section{Pole Model
for $D\ra\pi$ and  $B\ra\pi$ form factors}

The couplings $g_{D^*D\pi}$ and $g_{B^*B\pi}$ fix the normalization
of the form factors of the heavy-to-light transitions $D\ra\pi$ and
$B\ra\pi$, respectively, in the pole-model description \cite{BLN}:
\be
f^+_D(p^2)= \frac{f_{D^*}g_{D^*D\pi}}{2m_{D^*}(1-p^2/m_{D^*}^2)}~.
\label{onepole}
\ee
An analogous expression holds for the form factor $f^+_B(p^2)$.

It is difficult to justify the pole model from first principles.
Generally, it is believed that the vector dominance approximation is
valid
at zero recoil, that is at $p^2\ra m_D^2$. Arguments based
on heavy quark symmetry suggest a somewhat larger region
of validity characterized by
$(m_D^2-p^2)/m_c \sim O(1 $GeV). However,  there are no convincing
arguments
in favour of this model to be valid also at small values of $p^2$
which are
most interesting from a practical point of view.
Nevertheless, using the
results presented in Sections 2.2 and 2.3 one observes
that not only the shape but
also the absolute normalization of the form factors at low $p^2$
appears to be
in rough agreement with the pole model. This is illustrated
in Fig. 2.
Quantitatively, at $p^2=0$ we find
$
f^+_D(0)_{SR}= 0.66,~~     f^+_D(0)_{PM}=0.75 ~ $,
and
$
f^+_B(0)_{SR}= 0.29,~~     f^+_B(0)_{PM}= 0.44 ~$.
In the regions $m_Q^2-p^2 > O(1$ GeV$^2$) with $Q=c$ and $b$,
respectively, the numerical agreement between the light-cone sum rule
and the pole model is better than 15\% for  $f^+_D$, and still within
50\% for  $f^+_B$. This finding is surprising. Even if the
contributions of several low-lying resonances in the $D^*$
($B^*$) channel may mimic the $p^2$ dependence of a single pole,
there is no reason for
the normalization to be mainly given by the coupling
$g_{D^*D\pi}$ ($g_{B^*B\pi}$) to a good ( rough) approximation.

Despite of the overall qualitative agreement in the
mass range of $D$ and $B$ mesons,
the light-cone sum rule and the pole-dominance model
differ markedly in the asymptotic dependence
of the form factors on the heavy mass. Focusing on $B$ mesons and
using the familiar scaling laws
$
f_B\sqrt{m_B}= \hat{f}_B$, $~f_{B^*}\sqrt{m_B} = \hat{f}_{B^*}$
and $~g_{B^*B\pi} = (2 m_B/f_\pi)\hat g $
which are expected to be valid at $m_b\to\infty$ modulo logarithmic
corrections, the pole model predicts
$
f^+_B(0)_{PM} \sim 1/\sqrt{m_B} ~,
$
whereas the light-cone sum rule (\ref{formSR}) yields
$
f^+_B(0)_{SR} \sim 1/{m_B^{3/2}}~.
$
The latter result rests on
the behaviour in QCD of the leading twist pion
wave function near the end point, that is  on $\vp_\pi(u)\sim 1-u$
at $u \ra 1$.

Since we see no theoretical justification for
extrapolating the pole model to the region $p^2=0$ we believe
the sum rule result.
The solution suggested by Fig. 2 is then to match
the two descriptions in the region of intermediate momentum transfer
$ p^2\simeq m_Q^2-O(1$GeV$^2)$.
Referring for a detailed discussion to ref. \cite{BBKR}
we
emphasize  that the light-cone sum rules seem to be generally
consistent with the heavy quark expansion. In particular,
the light-cone sum rule (\ref{SR0}) correctly reproduces the heavy
quark
mass dependence of the coupling $g_{B^*B\pi}$. Fitting our
predictions for $g_{B^*B\pi}$ and $g_{D^*D\pi}$ to the form
\be\label{1/m}
g_{B^*B\pi} = \frac{2 m_B}{f_\pi}\cdot \hat g
\Bigg[1+\frac{\Delta}{m_B}\Bigg]
\ee
and the analogous expression for $g_{D^*D\pi}$,
we find for the coupling $\hat g$ and the strength $\Delta$
of the $1/m_Q$ correction:
\be
\hat g =0.32\pm 0.02~,~~ \Delta =(0.7 \pm 0.1)~ GeV ~.
\label{fit}
\ee

\section{ Nonfactorizable effects in the decay $B \ra J/\psi K $ }

Nonleptonic two-body decays of heavy mesons are usually calculated
by factorizing the appropriate matrix element of the
weak Hamiltonian $H_W$ into a product ( or a sum of such products ) of a
form factor and a decay constant. However, as well known, naive
factorization fails.
In order to achieve agreement with experiment it is necessary to let the
Wilson coefficients $a_{1,2}$   emerging from the operator product
expansion of $H_W$ and multiplying the relevant weak matrix elements
deviate from the
values predicted in short-distance QCD.
Phenomenologically \cite{BSW}, $a_{1,2}$ are treated as free
parameters to be determined from experiment.

The decay $B \rightarrow J/\psi K $ provides an important example.
The relevant part of the weak effective Hamiltonian may be written as
\be
H_W= \frac{G}{\sqrt{2}}V_{cb}V^*_{cs}\{(c_2+\frac{c_1}3) O_2+2c_1
\tilde{O}_2\}~,
\label{H}
\ee
with the four quark operators
\be
O_2=(\bar{c}\Gamma^\rho c)(\bar{s}\Gamma_\rho b),\
\tilde{O}_2=(\bar{c}\Gamma^\rho \frac{\lambda^a}2c)(\bar{s}\Gamma_\rho
\frac{\lambda^a}2 b)
\label{o}
\ee
and $\Gamma_\rho = \gamma_\rho(1-\gamma_5)$.
In factorization approximation, the decay amplitude
is given by
\be
\langle J/\psi(p) K(q)\mid H_W\mid B(p+q) \rangle
= \sqrt{2}GV_{cb}V_{cs}^*a_2f_\psi f_K^+m_\psi(\epsilon^\psi  \cdot q)
\label{factoriz}
\ee
where $a_2=c_2+ \frac{c_1}3$,
$f_{\psi}$ is
the decay constant of the $J/\psi$,
$f^+_K$ is the
$B \ra K $ form factor at $p^2=m_\psi^2$, and  $\epsilon^\psi$ denotes
the $J/\psi$ polarization vector. From the short-distance value of $a_2$,
the branching ratio is estimated to be almost an order of magnitude
smaller than the experimental result \cite{KuhnR,CLEO}. On the other hand,
dropping the term proportional to $c_1/3$ in $a_2$
as suggested in the framework of the $1/N_c$ expansion \cite{BGR}
of the weak amplitudes yields reasonable agreement.
One can argue that the factorizable term proportional
to $c_1/3$ is cancelled by nonfactorizable contributions being of the
same order in $1/N_c$.
Such a cancellation was first advocated in
ref. \cite{BGR} and then shown in ref. \cite{BS} to actually
take place in two-body $D$ decays . In the latter work QCD sum rule
techniques were used in order to estimate the nonfactorizable
amplitudes.

Recently, we have investigated
the problem of factorization in $B$ decays using
$B \rightarrow J/\psi K $ as a study case \cite{KLR}.
Following the general idea put forward in ref. \cite{BS},
we calculate the four-point correlation function
\be
<0\mid T\{j_{\mu5}^K(x)j_\nu^\psi(y)H_W(z)j^B_5(0)\}\mid 0>
\label{corr}
\ee
by means of the short-distance OPE.
Here $ j_{\mu5}^K= \bar{u}\gamma_\mu \gamma_5s $ ,
$ j_\nu^\psi= \bar{c}\gamma_\nu c $ and
$ j^B_5= \bar{b}i\gamma_5 u $
are the generating currents of the mesons involved and
$H_W$ is the effective weak Hamiltonian (\ref{H}).
To lowest nonvanishing order in $\alpha_s$ the
nonfactorizable contributions to the matrix element (\ref{corr})
only arise from the
operator $\tilde{O_2}$ in $H_W$.
Obviously, the contribution of this
operator to the $B \rightarrow J/\psi K $
amplitude (\ref{factoriz}) vanishes
by factorization because of colour conservation.
Parametrizing the nonfactorizable
matrix element by
\be
\langle J/\psi(p) K(q)\mid \tilde{O}_2 \mid B(p+q)\rangle =
2\tilde{f}f_\psi m_\psi(\epsilon^\psi \cdot q)
\label{nf}
\ee
we construct a sum rule
for $\tilde{f}$ which enters the correlation function
(\ref{corr}) through the ground state
contribution. In the QCD part of this sum rule all nonperturbative
contributions from vacuum condensates up
to dimension 6 are included. The corresponding diagrams are indicated in
Fig. 3.
In the hadronic part a complication arises from
intermediate states in the $B-$meson channel carrying the quantum
numbers of a $\bar{D}D_s^*$ pair. These virtual states are
created by weak interaction and converted into the
$J/\psi K $ final state by strong interaction.
In the quark-gluon representation  of the correlation function
(\ref{corr}) calculated from the diagrams of Fig. 3
one can identify corresponding four-quark
$\overline{u}s\overline{c}c$ intermediate states.
Invoking quark-hadron duality we cancel
this piece of the QCD part against the unwanted hadronic
contribution.

\vspace*{3.0cm}
\begin{center}
\begin{minipage}[t]{7.8cm} {
\begin{center}
\hspace{-3.2cm}
\mbox{
\epsfysize=10cm
\epsffile[0 0 500 500]{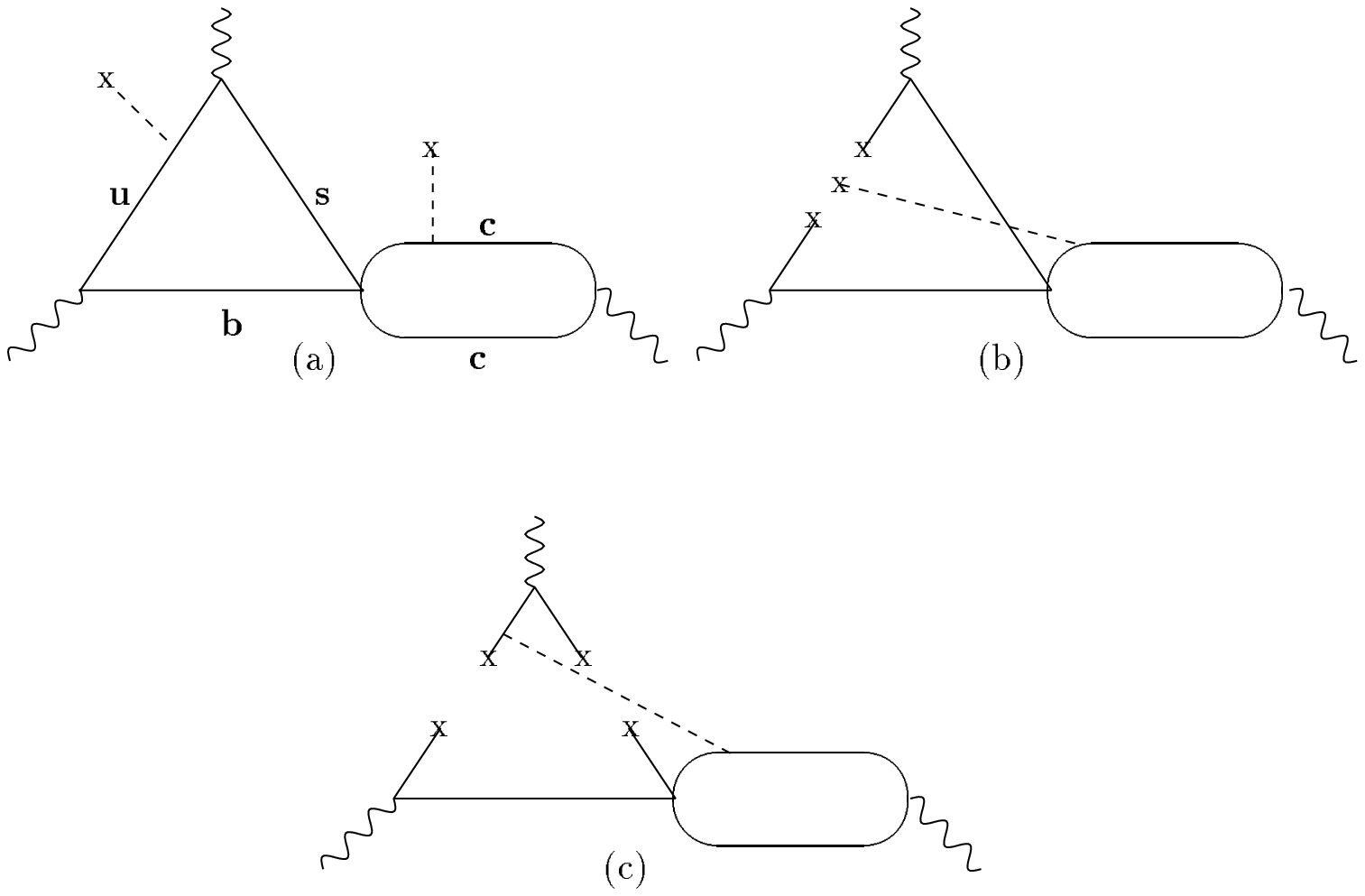}
}
\end{center}
%
}\end{minipage}
\end{center}
\vspace{-5.0cm}
{\small\bf Fig.~3. }{\small\it
Diagrams associated with (a) the gluon condensate,
(b) the quark-gluon condensate and (c) the four-quark condensate
contributions to the correlation function (27) with $H_W$
replaced by $\tilde{O}_2$.}
\bigskip

We then perform, as usual, a
Borel transformation in the $B-$meson channel and take
moments in the charmonium channel.
The spacelike momentum squared in the $K$-meson channel is
kept fixed. As explained in ref. \cite{KLR},
at this stage one encounters a second problem. The usual
subtraction of higher state contributions  employing semilocal
quark-hadron duality
is not possible here. Therefore, we must include these
contributions explicitly in the sum rule.
For this purpose we use a simple two-resonance model for the
spectral functions
in each of the three channels: $B$ and $B'$ in the $\bar{u}b$-channel,
$J/\psi$ and $\psi'$ in the $\bar{c}c$-channel,
and $K$ and $K'$ in the $\bar{u}s$-channel.
This rough approximation, yields
$\tilde{f} = -(0.045 \div 0.075)~$.
\label{ftilde}
The full decay amplitude for $B\ra J/\psi K $
is proportional to
\be
a_2=c_2+\frac{c_1}3 + 2c_1\tilde{f}/f_K^+~,
\label{a2}
\ee
where the first two coefficients are associated with the
factorizable part of the matrix element (\ref{factoriz}),
while the third term is due to the leading nonfactorizable
term (\ref{nf}). Interestingly enough, we find that the
factorizable nonleading in $1/N_c$
term $c_1/3$ and the nonfactorizable term in (\ref{a2})
are opposite in sign. Although the nonfactorizable matrix
element is considerably smaller than the factorizable one,
$|\tilde{f}/f^+_K|\simeq 0.1 $, it has a strong quantitative impact
due to its large coefficient,
$|2c_1/(c_2+c_1/3)| \simeq 20 \div 30 $. In fact, if
$|\tilde{f}|$ is close to the upper end of the predicted
range, the third term in eq. (\ref{a2}) almost cancels
the second term, thereby increasing the branching ratio
considerably. This is
exactly the scenario anticipated by $1/N_c$-rule \cite{BGR}.

It is also very interesting to note that our theoretical estimate yields
a negative overall sign for $a_2$ in contradiction to a global fit
to data \cite{CLEO}.
Furthermore, there is no theoretical reason in our approach
to expect universal values or even
universal signs for the coefficients $a_{1,2}$ in different channels,
in contrast to what seems to be suggested by experiment.
Universality can
at most be expected for certain classes of decay modes, such as
$B\rightarrow D\pi$ or $B \rightarrow D\overline{D}$, etc.
Also, there is no simple relation between $B$ and $D$ decays
in our approach since the  OPE for the corresponding correlation
functions significantly differ in the relevant diagrams and in the
hierarchy of mass scales.
We hope to be able to clarify these issues further.

Concluding we would like to stress that QCD
seems to predict a much richer pattern in two-body weak
decays than what is revealed by the current phenomenological
analysis of the data.

\section{Conclusion}
The flexible and careful employment of QCD sum rule techniques
in the analysis of exclusive heavy meson decays promises considerable
progress in solving the open problems, at least some of them.

\section{Acknowledgements}
We thank V. Braun, V. Belyaev and B. Lampe
for collaboration on the topics discussed in this report.
A. K. is grateful to the Alexander von Humboldt Foundation for
financial support during the initial stage of this work.


\end{document}